\RequirePackage{amsmath}
\documentclass[12pt]{iopart}

\usepackage{bm}
\usepackage{graphicx}
\usepackage{amsmath}
\usepackage{amsfonts}
\usepackage{bbold}
\usepackage{amssymb}
\usepackage{siunitx}
\usepackage{color}
\usepackage{cite}
\usepackage{comment}

\usepackage{hyperref}
\newcommand{\zt}{\bar{z}}
\newcommand{\Et}{\bar{E}}

\newcommand{\Ga}{\Gamma}
\newcommand{\abs}[1]{{\lvert}#1{\rvert}}%
\DeclareMathOperator{\IM}{Im}

\newcommand{\Gav}{\langle \widehat{G}(z)\rangle_\mathrm{av}}

\begin{document}

\title{Many-impurity scattering on the surface of a topological insulator}

\author{Jos\'{e} Luis Hernando, Yuriko Baba$^{*}$, Elena D\'{\i}az and Francisco Dom\'{i}nguez-Adame}

\address{GISC, Departamento de F\'{\i}sica de Materiales, Universidad Complutense, E--28040 Madrid, Spain}
\address{$^{*}$ Author to whom any correspondence should be addressed.}
\ead{yuribaba@ucm.es}

\begin{abstract}

We theoretically address the impact of a random distribution of non-magnetic impurities on the surface states formed at the interface between a trivial and a topological insulator. The interaction of electrons with the impurities is accounted for by a separable pseudo-potential method that allows us to obtain closed expressions for the density of states. Spectral properties of surface states are assessed by means of the Green's function averaged over disorder realizations. For comparison purposes, the configurationally averaged Green's function is calculated by means of two different self-consistent methods, namely the self-consistent Born approximation (SCBA) and the coherent potential approximation (CPA). The latter is often regarded as the best single-site theory for the study of the spectral properties of disordered systems. However, although a large number of works employ the SCBA for the analysis of many-impurity scattering on the surface of a topological insulator, CPA studies of the same problem are scarce in the literature. In this work we find that the SCBA overestimates the impact of the random distribution of impurities on the spectral properties of surface states compared to the CPA predictions. The difference is more pronounced when increasing the magnitude of the disorder.  

\end{abstract}

\pacs{       
 73.20.At,   
 73.22.Dj,   
 81.05.Hd    
}

\vspace{2pc}

\noindent{\it Keywords}: Topological insulators, disordered systems.


\maketitle

\section{Introduction}

Since the pioneering work of Anderson on the absence of diffusion in random lattices~\cite{Anderson58}, different models of disorder have played a major role in understanding optical and transport properties of real solids with point defects. The advent of two-dimensional (2D) Dirac materials, such as the surface of topological insulators, graphene and carbon nanotubes, has brought renewed interest in low-dimensional disordered systems. One of the most salient features of Dirac materials is the appearance of a gapless energy spectrum that depends linearly on momentum (Dirac cones). This dispersion makes electrons behave as massless fermions with a Fermi velocity much lower than the speed of light. The single-parameter hypothesis of disordered systems, introduced by Abrahams \emph{et al}.~\cite{Abrahams79}, led  to the general belief that all electron states were exponentially localized in 2D systems. Although this prediction works nicely when the energy spectrum depends quadratically on momentum, it turns out that extended states may arise in 2D systems with linear dispersion and quasi-particles undergo a localization-delocalization transition by varying the magnitude of disorder~\cite{Rodriguez03}. Therefore, it becomes apparent that electron dynamics in disordered 2D Dirac materials may substantially differ from what is known in conventional solids.
 
Single-particle spectral properties of disordered systems, such as the density of states (DOS), can be assessed by means of the Green's function averaged over disorder realizations~\cite{Gonis92,Economou06}. In general, the configurationally averaged Green's function cannot be calculated exactly and various approximations of different degree of sophistication are employed. Among them, self-consistent methods stand out because they correctly explain the main features of the DOS as inferred from photo-emission and soft X-ray experiments~\cite{Jones85}.  

Impurities and other point defects are common sources of disorder in 2D Dirac materials~\cite{Liu09,Xu17,Miao18,Dehnavi20}. Electron scattering by impurities yields spectral features, such as circular $s$-wave resonances, that can be targeted by scanning tunneling experiments (see reference~\cite{Xu17} and references therein). Theoretical treatments of many impurities are often based on the self-consistent Born approximation (SCBA)~\cite{Groth09,Fukuzawa09,Juan10,Noro10,Pesin15,Sbierski17,Sriluckshmy18,Klier19,Kudla19}. If the impurity potential is assumed short ranged, the SCBA leads to particularly simple expressions for the average Green's function, from which the DOS is readily determined. The so-called coherent potential approximation (CPA) represents another example of self-consistent approach routinely used for the theoretical analysis of conventional disordered matter~\cite{Gonis92,Weinberger90}. However, CPA studies of spectral and transport properties of disordered 2D Dirac materials are still scarce in the literature~\cite{Stauber08,Repetsky20a,Repetsky20b}, particularly in the context of surface states of topological insulators.

In this work we study many-impurity electron scattering on the surface of a topological insulator by means of self-consistent methods, namely SCBA and CPA, with the aim of comparing their predictions. The analysis and conclusions can be trivially extended to any 2D material where electron dynamics can be described by the massless Dirac equation. The interaction of the electron with the scatterers is accounted for by a separable pseudo-potential model~\cite{Knight63,Sievert73,Glasser75,Adame91,Adame95,Prunele97,Lopez02,Gonzalez_Santander13}. In spite of its seemingly more complicated form, the separable pseudo-potential model is amenable to analytical solution and allows us to obtain closed expressions for the average Green's function within the SCBA and CPA frameworks. In particular, short-range potentials approaching the $\delta$-function limit, frequently used in previous works~\cite{Fukuzawa09,Klier19,Lu11,Biswas10,Shiranzaei17}, can be viewed as limiting cases of the separable pseudo-potential model. We will show that the SCBA average Green's function can also be obtained from the CPA calculations in the limit of diluted impurities and small magnitude of disorder. However, the main conclusion of this work is that the SCBA overestimates the impact of point-like scatterers on the spectral properties, compared to the CPA predictions. The discrepancy becomes greater when increasing the impurity concentration. 

\section{Theoretical model} \label{sec:theory}

The Hamiltonian operator of an electron in a pristine surface of a topological insulator will be denoted as $\widehat{H}_0$. It is diagonal in the basis of plane waves
\begin{subequations}
\begin{equation}
  \langle\, {\bm k} \mid \widehat{H}_0 \mid {\bm k}^{\prime}\,\rangle = H_{0}({\bm k})\, \delta_{{\bm k},{\bm k}^{\prime}}\ ,
  \label{eq:01a}
\end{equation}
where~\cite{Ortmann15}
\begin{equation}
  H_{0}({\bm k})=\hbar v \left(\sigma_x k_y - \sigma_y k_x\right)\ .
  \label{eq:01b}
\end{equation}
\label{eq:01}%
\end{subequations}
Here $v$ is a matrix element having dimensions of velocity, $\sigma_x$ and $\sigma_y$ are Pauli matrices and ${\bm k}=(k_x,k_y)$ is the in-plane momentum. The corresponding bands are simply given as $E_{\bm k}=\pm \hbar v|{\bm k}|$ (Dirac cones). 

Let us address how the electron interacts with impurities located at the surface of the topological insulator. We will assume that they are placed on a regular square lattice of parameter $a$. Notice that $a$ is not related to the size of the unit cell of the crystal structure of the topological insulator. In fact, electrons do not \emph{see\/} the crystal structure since we are using a continuous approximation for the Hamiltonian~(\ref{eq:01}). We will focus on binary disorder hereafter. To this end, two different species of impurities A and B are considered. A given site of the square lattice is occupied by an impurity A with probability $c$  or by an impurity B with probability $1-c$. Therefore, the separable pseudo-potential operator can be cast in the form~\cite{Sievert73,Prunele97}
\begin{equation}
  \widehat{V}=\sum_n  \widehat{V}_n\ ,
  \qquad
  \widehat{V}_n = \mid\omega_n \,\rangle \lambda_n \langle\, \omega_n\mid \ .
  \label{eq:02}
\end{equation}
The index $n$ runs over all sites ${\bm R}_n$ of the square lattice and $\omega({\bm r}-{\bm R}_n)=\langle\,{\bm r}\mid \omega_n\,\rangle$ will be referred to as \emph{shape function}. $\lambda_n$ is the coupling constant that takes on two values $\lambda_\mathrm{A}$ and $\lambda_\mathrm{B}$ at random, with probability $c$ and $1-c$ respectively. Hence, the probability distribution in this model of binary disorder is
\begin{equation}
  \mathcal{P}(\lambda_n)=c\delta(\lambda_n-\lambda_\mathrm{A})+(1-c)\delta(\lambda_n-\lambda_\mathrm{B})\ .
  \label{eq:03}
\end{equation}

The electron Hamiltonian in the presence of the impurities is the sum of the Hamiltonian $\widehat{H}_0$ corresponding to the translationally invariant system and the random part $\widehat{V}$, namely $\widehat{H}=\widehat{H}_0+\widehat{V}$. The retarded Green's function operators (resolvents) corresponding to $\widehat{H}$ and $\widehat{H}_0$ are 
\begin{equation}
  \widehat{G}(z)=\left(z-\widehat{H}\right)^{-1}\ ,
  \qquad
  \widehat{G}_0(z)=\left(z-\widehat{H}_0\right)^{-1}\ ,
  \label{eq:04}
\end{equation}
where $z=E+i0^{+}$. Notice that $\widehat{G}_0(z)$ is diagonal in the basis of plane waves
\begin{subequations}
\begin{equation}
  \langle\, {\bm k} \mid \widehat{G}_0(z) \mid {\bm k}^{\prime}\,\rangle = G_{0}({\bm k},z)\, \delta_{{\bm k},{\bm k}^{\prime}}\ ,
  \label{eq:05a}
\end{equation}
with
\begin{equation}
  G_{0}({\bm k},z)=\frac{1}{z-H_0({\bm k})}=\frac{z+H_0({\bm k})}{z^2-\hbar^2v^2k^2}\ ,
  \label{eq:05b}
\end{equation}
\label{eq:05}%
\end{subequations}
by virtue of equation~(\ref{eq:01}). We will concern ourselves with the ensemble average $ \Gav$ of the Green's function operator in the random medium. The subscript `$\mathrm{av}$' indicates the average over the probability distribution~(\ref{eq:03}).

The knowledge of $\Gav$ allows us to obtain the spectral properties of an electron on the surface of a topological insulator scattered off by a random array of impurities. In general, the average Green's function operator cannot be obtained exactly and some approximations are needed. The conceptually simplest way of finding an approximation to $\Gav$ is by introducing an effective, translationally invariant medium represented by a Green's function operator $\widehat{G}_\mathrm{eff}(z)$ such that $\widehat{G}_\mathrm{eff}(z)=\Gav$. The first level of approximation is reached in the case of very weak scattering by assuming that the array of impurities is periodic with a coupling constant given as the following average
\begin{equation}
  \lambda_\mathrm{VCA}\equiv \langle \lambda_n \rangle_\mathrm{av}=c \lambda_\mathrm{A}+(1-c)\lambda_\mathrm{B}\ .
  \label{eq:06}
\end{equation}
This approach is known as the Virtual Crystal Approximation (VCA) (see, e.g., reference~\cite{Gonis92}). The VCA is a reasonably good description only if $c\to 0$ (or equivalently $c\to 1$) and $\lambda_\mathrm{A}\simeq \lambda_\mathrm{B}$. However, more elaborated, self-consistent methods have a much wider range of validity. Within these methods, the VCA appears usually as a first constant term in the expansion of the Green's function, as described in the following sections. 

Once the effective Green's function is obtained, important physical quantities can be calculated. In particular, the average DOS per unit area is easily computed by the following expression
\begin{equation} \label{eq:rho:def}
    \rho(E) = -\frac{1}{\pi S}\IM \Big[\Tr\big(\hat{G}_\mathrm{eff} (E+i0^+)\big)\Big]~.
\end{equation}
We will take $S=1$ and referred to $\rho(E)$ as the DOS hereafter.

\section{Self-consistent Born approximation} \label{sec:scba}

In this section we consider the effects of disorder within the SCBA. In the framework of this approximation, the Green's function operator of the effective medium is taken as 
\begin{subequations}
\begin{equation}
  \widehat{G}_\mathrm{eff}(z) = \widehat{G}_0\left[z-\widehat{\Sigma}_\mathrm{SCBA}(z)\right]\ ,
  \label{eq:07a}
\end{equation}
where the self-energy operator $\widehat{\Sigma}_\mathrm{SCBA}(z)$ is diagonal in the basis of plane waves 
\begin{equation}
  \langle\, {\bm k} \mid \widehat{\Sigma}_\mathrm{SCBA}(z) \mid {\bm k}^{\prime}\,\rangle 
  = \Sigma_\mathrm{SCBA}({\bm k},z)\, \delta_{{\bm k},{\bm k}^{\prime}}\ .
  \label{eq:07b}
\end{equation}
The self-energy $\Sigma_\mathrm{SCBA}({\bm k},z)$ is to be determined self-consistently from the following equation
\begin{align}
  \Sigma_\mathrm{SCBA}({\bm k},z)&=\frac{|{\omega (\bm{k})}|^2}{a^2} \lambda_{\mathrm{VCA}}\nonumber \\
  &+ \int \frac{\mathrm{d}^2{\bm k}^{\prime}}{4\pi^2}\,C({\bm k}-{\bm k}^{\prime}) \, 
  G_0\big[{\bm k}^{\prime},z-\Sigma_\mathrm{SCBA}({\bm k}^{\prime},z)\big]\ .
  \label{eq:07c} 
\end{align}
\label{eq:07}%
\end{subequations}
Here $C({\bm k}-{\bm k}^{\prime})$ is the disorder correlator that depends on the transferred momentum. In the case of the separable pseudo-potential model~(\ref{eq:02}) we get
\begin{equation}
  C({\bm k}-{\bm k}^{\prime})=\frac{1}{a^2}|\omega({\bm k})\,
  \omega({\bm k}^{\prime})|^2 c  \Delta^2\ ,
  \label{eq:08}
\end{equation}
where $\omega({\bm k})=\int \mathrm{d}^2{\bm r}\,e^{i{\bm k}\cdot{\bm r}}\omega({\bm r}) $ is the Fourier transform of the shape function and $\Delta=\lambda_\mathrm{A}-\lambda_\mathrm{B}$ is the magnitude of disorder.

For convenience, we define the self-energy as the product of an effective  coupling  constant $\lambda_\mathrm{SCBA}(z)$ and the shape function $\omega({\bm k})$ as follows
\begin{subequations}
\begin{equation}
  \Sigma_\mathrm{SCBA}({\bm k},z) = \lambda_\mathrm{SCBA}(z)\,\frac{|\omega({\bm k})|^{2}}{a^2}\ .
  \label{eq:09a}
\end{equation}
Therefore, equation~(\ref{eq:07c}) is written as
\begin{align}
  \lambda_\mathrm{SCBA}(z) & = \lambda_\mathrm{VCA} + 
  c \Delta^2 \int \frac{\mathrm{d}^2{\bm k}}{4\pi^2}\, |\omega({\bm k})|^2 \, G_0\big[{\bm k},z-\lambda_\mathrm{SCBA}(z)
  |\omega({\bm k})|^2/a^2\big]\ .
  \label{eq:09b}
\end{align}
\label{eq:09}%
\end{subequations}

Notice that equation~(\ref{eq:09b}) is valid for any shape function and consequently it is suitable for the study of finite-range impurity potentials. However, particularly simple expressions are found for point-like impurities, namely when $\omega({\bm k})$ becomes independent of ${\bm k}$. Since the resulting integral is divergent at large momenta, we impose a momentum cutoff $k_\mathrm{c}$ (or, equivalently, we introduce a finite bandwidth) and set
\begin{equation}
  \omega({\bm k})=\omega(k)=a\theta(k_\mathrm{c}-k)\ ,
  \label{eq:10}
\end{equation}
where $\theta$ is the Heaviside step function and the impurity lattice constant $a$ is introduced for convenience. Notice that $H_0({\bm k})$ in the numerator of~(\ref{eq:05b}) is an odd function of momentum and consequently the corresponding integration vanishes.  Moreover, we find it more convenient to express the results in terms of the coupling constant obtained within the VCA \eqref{eq:06} by defining $\Lambda_\mathrm{SCBA}(\bar{z}) =\lambda_\mathrm{SCBA}(z)-\lambda_\mathrm{VCA}(z)$ with $\bar{z}=z-\lambda_\mathrm{VCA}(z)$. Hence, the self-consistent equation for SCBA can be expressed in a compact form as follows
\begin{subequations}
\begin{equation}
  \frac{\Lambda_\mathrm{SCBA}(\bar{z})}{c\Delta^2}
  =\mathcal{F}\left[\bar{z}-\Lambda_\mathrm{SCBA}(\bar{z})\right] \ ,
  \label{eq:selfSCBA}
\end{equation}
where we have defined
\begin{equation}
  \mathcal{F}(z)=\frac{a^2}{4\pi}\int_{0}^{k_\mathrm{c}} \mathrm{d}k\,k\left(\frac{1}{z+\hbar v k}+\frac{1}{z-\hbar v k}\right)\ .
  \label{eq:11b}
\end{equation}
\label{eq:11}%
\end{subequations}

\section{Coherent potential approximation} \label{sec:cpa}

The CPA traces back to the sixties and has proven to be a successful mean field theory for the study of various elementary  excitations (electrons, phonons, excitons, magnons) in disordered systems~\cite{Soven67,Taylor67,Onodera68,Velicky69}. The CPA combines two basic ideas. On one side, the average Green's function of the disordered system is calculated by introducing a periodic (translationally invariant) effective medium. On the other hand, this effective medium is determined by demanding that the fluctuations of the Green's function average out to zero, thus leading to a self-consistency condition~\cite{Economou06}. In the single-site CPA combined with the separable pseudo-potential model, the electron motion in the effective medium is represented by the following Hamiltonian~\cite{Sievert73,Glasser75}
\begin{equation}
  \widehat{H}_\mathrm{eff}=\widehat{H}_0+\sum_{n} \mid\omega_n \,\rangle \lambda_\mathrm{CPA}(z) \langle\, \omega_n\mid\ ,
  \label{eq:12}
\end{equation}
where $\lambda_\mathrm{CPA}(z)$ is in general complex and will be determined self-consistently from the condition $\widehat{G}_\mathrm{eff}(z) = (z-\widehat{H}_\mathrm{eff})^{-1} = \langle \widehat{G}(z)\rangle_\mathrm{av}$. In contrast to $\widehat{H}=\widehat{H}_0+\widehat{V}$ with $\widehat{V}$ given by equation~(\ref{eq:02}), the effective Hamiltonian $\widehat{H}_\mathrm{eff}$ has the full symmetry of the impurity lattice since $\lambda_\mathrm{CPA}(z)$ is taken to be independent of the site. The difference between both Hamiltonians can be expressed as
$\widehat{H} - \widehat{H}_\mathrm{eff} = \sum_{n}\widetilde{V}_n$ with
\begin{equation}
  \widetilde{V}_n=\mid\omega_n \,\rangle \left[\lambda_n-\lambda_\mathrm{CPA}(z)\right] \langle\, \omega_n\mid \ .
  \label{eq:13}
\end{equation}

To proceed, we consider the $t$-matrix operator associated with a single site~\cite{Gonis92}
\begin{align}
  \widehat{t}_n(z)&=\left[1-\widetilde{V}_n\widehat{G}_\mathrm{eff}(z)\right]^{-1}\widetilde{V}_n 
   =\sum_{m=0}^{\infty}\left[\widetilde{V}_n\widehat{G}_\mathrm{eff}(z)\right]^m\widetilde{V}_n \nonumber \\
  & = \frac{\mid \omega_n\,\rangle \left[\lambda_n-\lambda_\mathrm{CPA}(z)\right] \langle\, \omega_n\mid}%
  {1-\left[\lambda_n-\lambda_\mathrm{CPA}(z)\right]\langle\, \omega_n \mid \widehat{G}_\mathrm{eff}(z)\mid\omega_n\,\rangle}\ .
  \label{eq:14}
\end{align}
It can be proven that the requirement $\widehat{G}_\mathrm{eff}(z) = \langle \widehat{G}(z)\rangle_\mathrm{av}$ yields the well-known CPA condition~\cite{Gonis92,Jones85,Economou06}
\begin{equation}
  \left\langle\widehat{t}_n(z)\right\rangle_\mathrm{av} = 0\ . \medskip
  \label{eq:CPA:cond}
\end{equation}
Therefore, from~(\ref{eq:14}) we finally get
\begin{subequations}
\begin{equation}
  \left\langle\frac{\lambda_n-\lambda_\mathrm{CPA}(z)}%
  {1-\left[\lambda_n-\lambda_\mathrm{CPA}(z)\right]\langle\, \omega_n \mid \widehat{G}_\mathrm{eff}(z)\mid\omega_n\,\rangle} \right\rangle_\mathrm{av} = 0\ ,
  \label{eq:16a}
\end{equation}
where, within the one-band approximation (see \ref{app:one-band}), we have
\begin{gather}
  \langle\, \omega_n \mid \widehat{G}_\mathrm{eff}(z)\mid\omega_n\,\rangle = \int \frac{\mathrm{d}^2{\bm k}}{4\pi^2} |\omega({\bm k})|^2 \, G_0\big[{\bm k},z-\lambda_\mathrm{CPA}(z)
   |\omega({\bm k})|^2/a^2\big]\ .
  \label{eq:16b}
\end{gather}
It is worth mentioning that $\langle\, \omega_n \mid \widehat{G}_\mathrm{eff}(z)\mid\omega_n\,\rangle$ becomes site independent since the effective medium in translationally invariant. Thus, the ensemble average in the case of binary disorder~(\ref{eq:03}) poses no problem and~(\ref{eq:16a}) leads to
\begin{gather}
  \frac{c}{\lambda_\mathrm{B}-\lambda_\mathrm{CPA}(z)}+ \frac{1-c}{\lambda_\mathrm{A}-\lambda_\mathrm{CPA}(z)} =
  \int \frac{\mathrm{d}^2{\bm k}}{4\pi^2} |\omega({\bm k})|^2 \, G_0\big[{\bm k},z-\lambda_\mathrm{CPA}(z)
  |\omega({\bm k})|^2/a^2\big]\ .
  \label{eq:16c}
\end{gather}
\label{eq:16}%
\end{subequations}

The above expression is valid for any shape function. In particular, in the case of point-like impurities~(\ref{eq:10}) one gets
\begin{equation}
  \frac{c}{\lambda_\mathrm{B}-\lambda_\mathrm{CPA}(z)}+ \frac{1-c}{\lambda_\mathrm{A}-\lambda_\mathrm{CPA}(z)}=
  \mathcal{F}\left[z-\lambda_\mathrm{CPA}(z)\right]\ ,
  \label{eq:17}
\end{equation}
where $\mathcal{F}(z)$ is defined in~(\ref{eq:11b}). Once more, it is more convenient to express the left-hand side of equation~(\ref{eq:17}) in terms of the coupling constant obtained within the VCA~(\ref{eq:06}) by defining
\begin{align}
  \Lambda_\mathrm{CPA}(\bar{z})& =\lambda_\mathrm{CPA}(z)-\lambda_\mathrm{VCA}(z)\ ,
\end{align}
whence
\begin{equation}
  \frac{\Lambda_\mathrm{CPA}(\bar{z})}{\left[c\Delta+\Lambda_\mathrm{CPA}(\bar{z})\right]\left[(1-c)\Delta-\Lambda_\mathrm{CPA}(\bar{z})\right]}
  =\mathcal{F}\left[\bar{z}-\Lambda_\mathrm{CPA}(\bar{z})\right]\ .
  \label{eq:selfCPA}
\end{equation}

\begin{figure}[htb]
    \centering
    \includegraphics[width=0.5\linewidth]{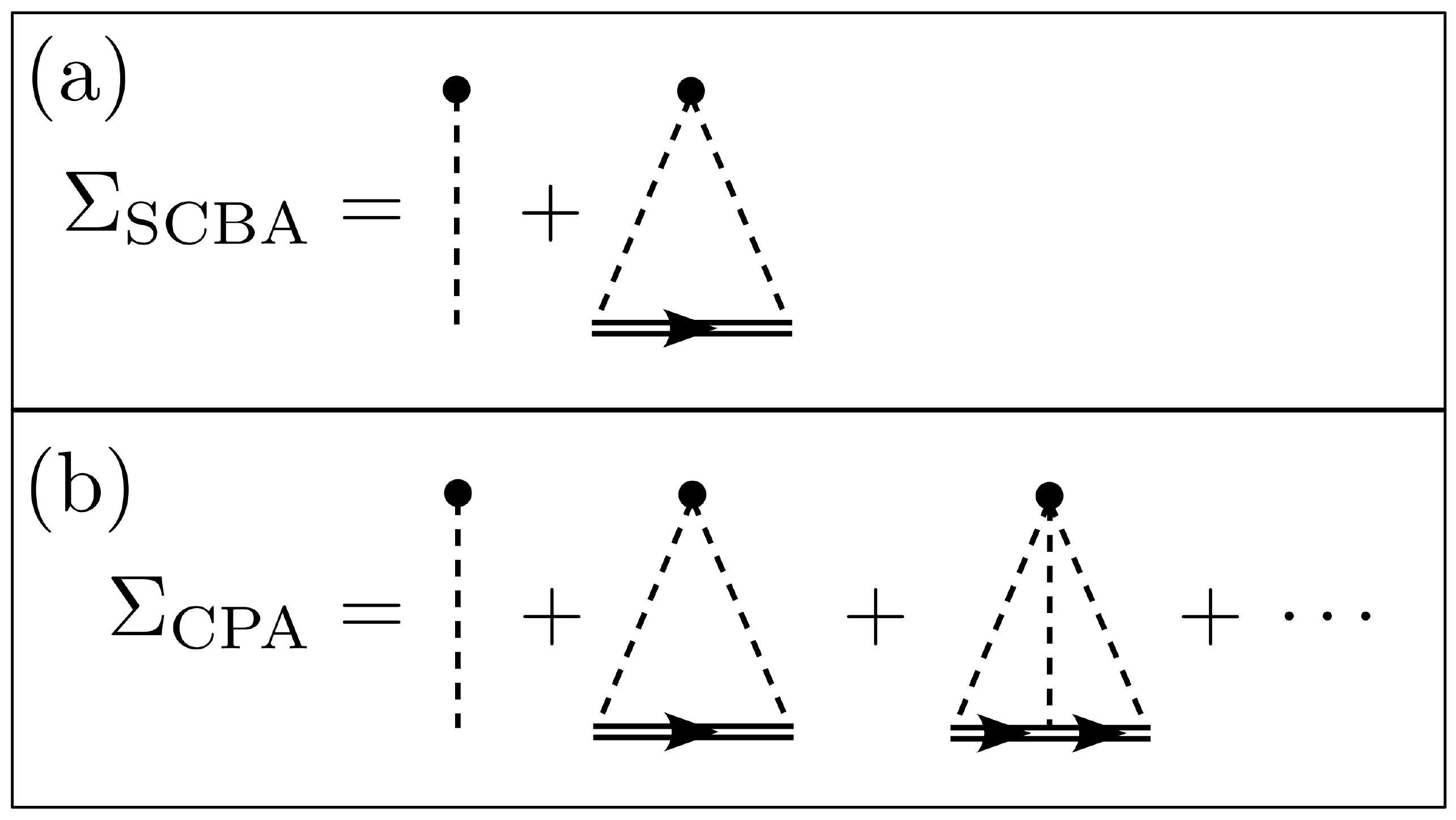}
    \caption{Irreducible diagrams that are taken into account in the calculation of the 
    self-energy in the (a)~SCBA and (b)~CPA. In the Feynman diagrams, the dashed line represents the scattering amplitude (i.e. the magnitude of the disorder), the double solid line is the effective propagator and the dot corresponds to the impurity (i.e. the vertex of the momentum-conserved interaction). 
    }
    \label{fig:Diagrams}
\end{figure}

Notice that, expanding the CPA self-consistent equation given by \eqref{eq:selfCPA}, we can get the SCBA. This can be obtained by solving for $\Lambda_{\mathrm{CPA}}(\bar{z})$ and expanding the result in a Taylor series for small $c$ and $\Delta$ up to third order
\begin{equation}
    \Lambda_\mathrm{CPA}(\bar{z})  = c \Delta^2 
    \mathcal{F}\left[\bar{z}-\Lambda_\mathrm{CPA}(\bar{z})\right] 
    + \mathcal{O}(c^2, \Delta^3)~ . 
    \label{eq:CPAseries}
\end{equation}
In fact, the SCBA can be obtained as a truncation of the series of the CPA. This is further clarified in the diagrammatic formalism with Feynman rules. The SCBA takes into account the two irreducible diagrams shown in figure~\ref{fig:Diagrams}(a) for the self-energy. The first diagram is the constant VCA term while the second one describes the double scattering off by a single impurity with a dressed internal propagator. On the other hand, CPA sums all the diagrams with any number of scattering events on the same impurity that, upon a proper re-summation~\cite{Elliot74}, gives the self-consistent equation \eqref{eq:selfCPA} [see figure~\ref{fig:Diagrams}(b)].

\section{Results} \label{sec:results}

In this section we analyse and compare the results obtained within both self-consistent approximations. First of all, we discuss the effective coupling constant, or equivalently the self-energy, obtained by SCBA and CPA. Later, we will present the results for the DOS. For simplicity, we cast the effective coupling constant as
\begin{equation} \label{eq:Lambda}
    \Lambda (\Et) = \alpha (\Et)  - i \Ga (\Et)~,
\end{equation}
where $\Lambda$ refers either to $\Lambda_\mathrm{SCBA}$ or $\Lambda_\mathrm{CPA}$, $\alpha(\bar{E})$ is real and $\Ga (\bar{E})> 0$ corresponds to a disorder-induced broadening. Notice that we express the results as a function of $\Et$ that is the shifted energy after taking the limit of $\zt \to  \Et + i0^+$.

By analyzing the symmetry properties of the CPA self-consistent equation~\eqref{eq:selfCPA}, we find that it is invariant under the exchange $(\alpha, \Ga, \Et, c)\to(-\alpha, \Ga, -\Et, 1-c)$ and  $(\Delta, c)\to(-\Delta, 1-c)$. Therefore, we can restrict ourselves to $\Delta>0$ and $0 \leq c < 0.5$ since all the other scenarios can be obtained from the former range of parameters. 
For the SCBA self-consistent equation~\eqref{eq:selfSCBA}, less symmetries are obtained due to the truncation of the series expansion of the CPA (see figure~\ref{fig:Diagrams}). Once high-order terms in $c$ are neglected ($c\to 0$), the symmetry $(\Delta, c)\to(-\Delta, 1-c)$ is lost. Notice that in order to investigate the range $c\geq 0.5$, the expansion in equation~\eqref{eq:CPAseries} must be performed around $1-c$ instead of $c$. Hence, the expression \eqref{eq:selfSCBA} can be used only for $0 \leq c<0.5$. 

Remarkably, apart from the constant VCA term, the SCBA depends on a single parameter related to disorder, namely  $c\Delta^2$, while the CPA needs both $c$ and $\Delta$ separately. We define the SCBA \emph{disorder parameter \/}$\beta$ as follows:
\begin{equation} \label{eq:beta:def}
    \beta\equiv \frac{c \Delta^2 a^2}{4\pi( \hbar v)^2}~,
\end{equation}
Hence, expressing the energies in units of $\hbar v /a$, we can write equation \eqref{eq:selfSCBA} as a function of a single dimensionless disorder parameter\/ given by equation~\eqref{eq:beta:def} and the energy cut-off $E_c\equiv \hbar v k_c$

\begin{equation} \label{eq:selfSCBA:beta}
    \Lambda_\mathrm{SCBA} = {\beta (\Et-\Lambda_\mathrm{SCBA})}  \ln\left[ \frac{(\Et- \Lambda_\mathrm{SCBA})^2}{(\Et-\Lambda_\mathrm{SCBA})^2-E_c^2}\right]~,
\end{equation}
where it is understood that $\Lambda_\mathrm{SCBA}=\Lambda_\mathrm{SCBA}(\Et)$. This equation is invariant under the exchange $(\alpha, \Ga, \Et) \to (-\alpha, \Ga, -\Et)$. Hence, the real part of the effective coupling constant is an odd function of the shifted energy, $\alpha(\Et) = -\alpha(-\Et)$, while the imaginary part is an even function, $\Ga(\Et) = \Ga(-\Et) $. With these considerations in mind, the SCBA equation \eqref{eq:selfSCBA:beta} can be solved explicitly in the case of $\Et = 0$, finding the zero-energy solution
\begin{equation}
    \alpha(\Et=0) = 0~,
    \qquad
    \Ga(\Et=0) = E_c\,\exp\left(\frac{1}{2}-\frac{1}{2\beta}\right)~.
    \label{eq:SCBA:G_E0}
\end{equation} 
This exponential behaviour is opposite to the case of single-node Weyl semimetals studied in reference~\cite{Klier19}, where a critical point signals a disorder-induced phase transition. In the Dirac-like Hamiltonian \eqref{eq:01b}, no critical behaviour is observed as a function of the magnitude of disorder, as discussed later.

\begin{figure}[htb]
    \centering
    \includegraphics[width=\linewidth]{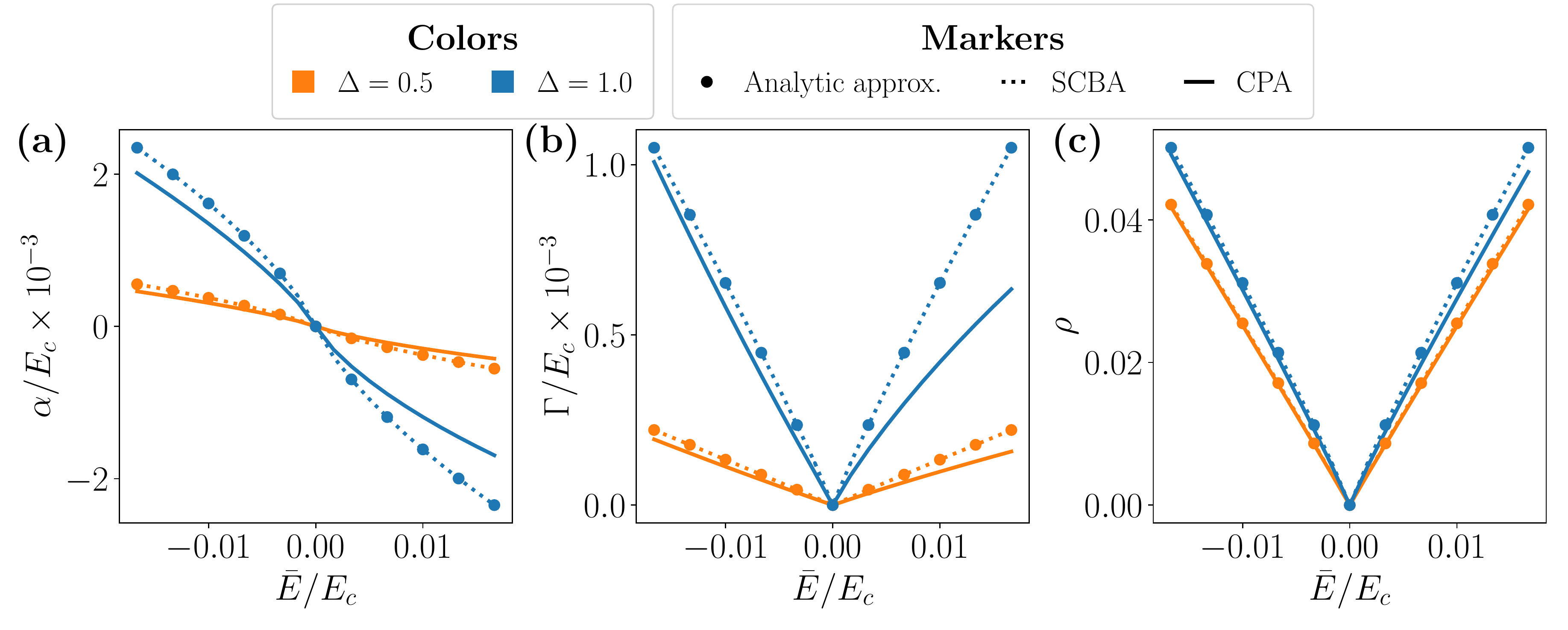}
    \caption{Coupling constant and DOS as a function of the shifted energy. (a)~Real part of the coupling constant $\alpha$, (b)~complex part of the coupling constant $\Gamma$ and (c) DOS $\rho$. The plots compare the results for fixed $c=0.2$ and two values of $\Delta = [0.5, 1.0]$ represented in orange and blue, respectively. Three approaches are compared: The analytic solution given by equation~\eqref{eq:Small:Lambda}, the SCBA and the CPA. $\Delta$ is expressed in units of $\hbar v/a$ and $E_c=15$.}
    \label{fig:AnalyticsComp}
\end{figure}

The SCBA self-consistent equation can be solved analytically for energies $\abs{\Et} \ll E_c$ and small disorder $\abs{\Lambda_\mathrm{SCBA}(\Et)} \ll E_c$ (see \ref{app:SCBA:analytics} for details). In this regime, the coupling constant can be approximated as
\begin{equation} \label{eq:Small:Lambda}
    \Lambda_\mathrm{SCBA} (\Et) \simeq
    \Et \left\{ 1- \frac{1}{2\beta}\left[ \mathcal{W}\left(
    -i \,\frac{ \Et  }{2 \beta  E_c}\,e^{1/(2\beta)}
    \right)\right]^{-1}\right\}~ ,
    \label{Lambert}
\end{equation}
where $\mathcal{W}(x)$ is the Lambert-W function~\cite{Corless96}. Figure \ref{fig:AnalyticsComp} shows a comparison of the analytic expression for the coupling constant with the numerically solved SCBA equation as a function of energy for weak disorder. Notice that the approximated solution agrees exceedingly well with the numerics, as long as the range of parameters considered fulfils all the conditions for the approximation to be valid. For the sake of completeness, the CPA results are plotted in solid lines as well. It is worth mentioning that the CPA and the SCBA results coincide very nicely for small disorder, as predicted by the series-expansion interpretation of the SCBA introduced in the previous section. Most importantly, upon increasing the magnitude of the disorder and the energy, the SCBA tends to overestimate the impact of the impurities on the DOS. 

In the following, we analyse the limit of $\Et \to 0$. As already mentioned, the SCBA predicts an exponential-like broadening given by equation \eqref{eq:SCBA:G_E0} that resembles, for small disorder parameter, a purely exponential decay [see \ref{app:SCBA:analytics} for further details]
\begin{equation} \label{eq:small:G_E0}
    \Gamma_\mathrm{SCBA} (\Et=0)  \simeq E_c\, e^{-1/(2\beta)}~. 
    \label{expansion-Lambert}
\end{equation}
The above expression of the broadening allows us to write explicitly the value of the DOS in the zero-energy limit within the SCBA [see \ref{app:DOS} for further details]
\begin{equation}
    \rho_{\mathrm{SCBA}} (\Et=0) 
     = \frac{E_c}{2 \pi^2 \beta \sqrt{e^{1/\beta } -1}} 
     \simeq  \frac{E_c}{2 \pi^2 \beta}\, e^{- 1/(2 \beta)}~. \label{eq:small:rho_E0}    
\end{equation}
\begin{figure}[htb]
    \centering
    \includegraphics[width=\linewidth]{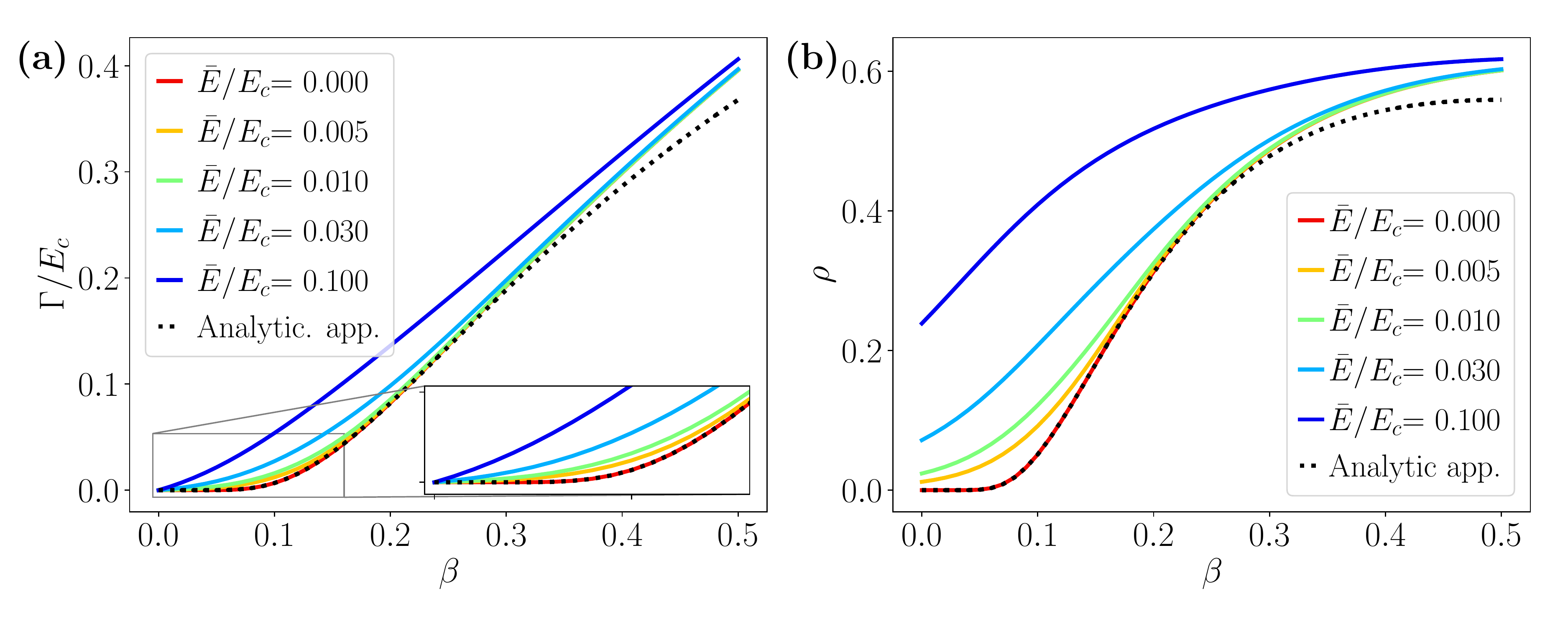}
    \caption{Imaginary part of the (a)~effective coupling constant and (b)~DOS within the SCBA as a function of the disorder parameter $\beta$. Dotted lines correspond to equations~\eqref{eq:small:G_E0} and \eqref{eq:small:rho_E0} while solid lines show the results from self-consistent calculations. Energy is expressed in units of $\hbar v/a$ and $E_c=15$.}
    \label{fig:SCBA:beta}
\end{figure}
Figure \ref{fig:SCBA:beta} shows a comparison of the analytic limit and the results of the numerically-solved $\Gamma(\Et)$ and DOS as a function of the disorder parameter $\beta$ within the SCBA. The absence of a disorder-induced phase transition is patent. 
\begin{figure}[htb]
    \centering
    \includegraphics[width=\linewidth]{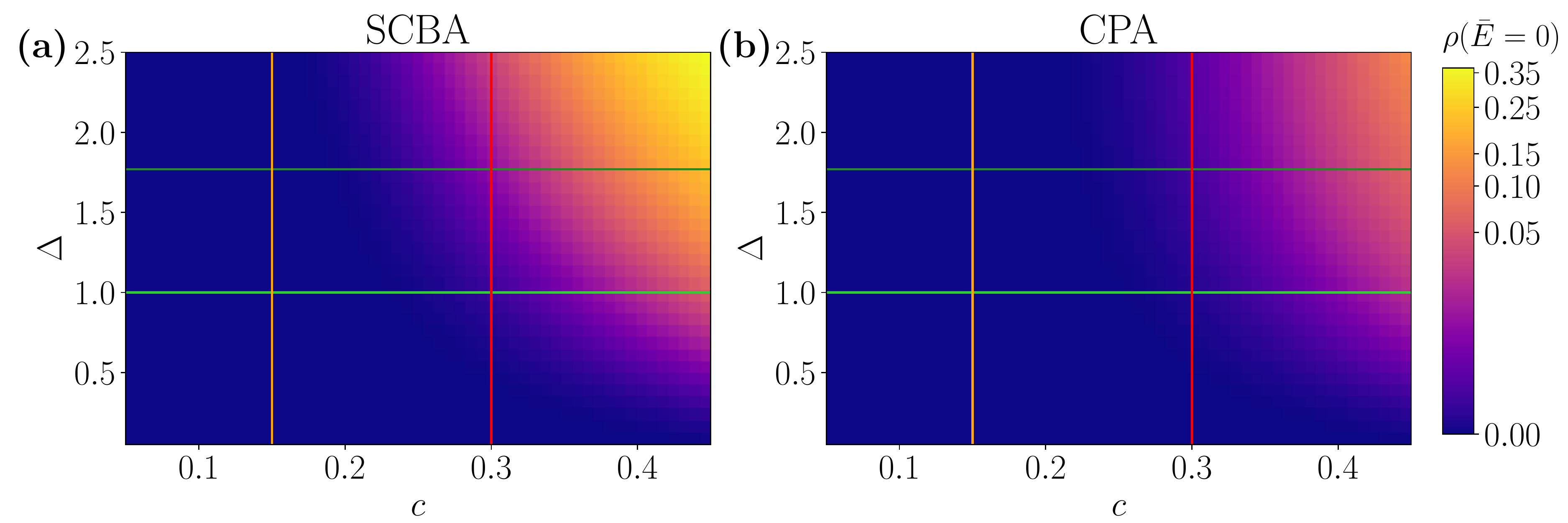}\\
    \includegraphics[width=\linewidth]{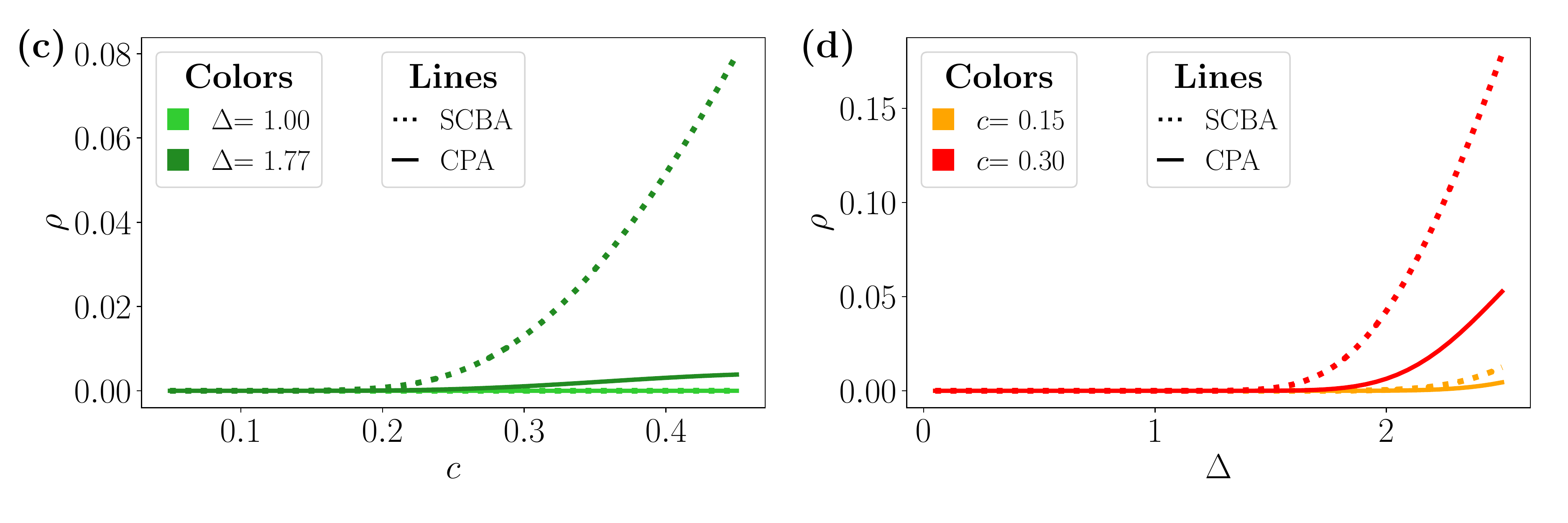}
    \caption{DOS at zero shifted energy as a function of the disorder strength $\Delta$ and the fraction $c$ of A impurities for both approximations. The cut-off energy is set to $E_c=15$, where energy and $\Delta$ are expressed in units of $\hbar v/a$.}
    \label{fig:EzeroPhases}
\end{figure}
The absence of a phase transition is observed in the CPA results as well. In fact, we find a smooth dependence of the DOS at zero energy $\Et$ on the disorder magnitude $\Delta$ and the fraction $c$ of A impurities, as seen in figure~\ref{fig:EzeroPhases}. Notice that in the CPA both parameters are needed and they can not be combined into a single disorder parameter, as we already found in the SCBA. The aforementioned figure reproduces again another important aspect of the predictions of both methods, namely the overestimation of the SCBA compared to the CPA. The disagreement becomes more marked when increasing the magnitude of disorder.

\begin{figure}
    \centering
    \includegraphics[width=\linewidth]{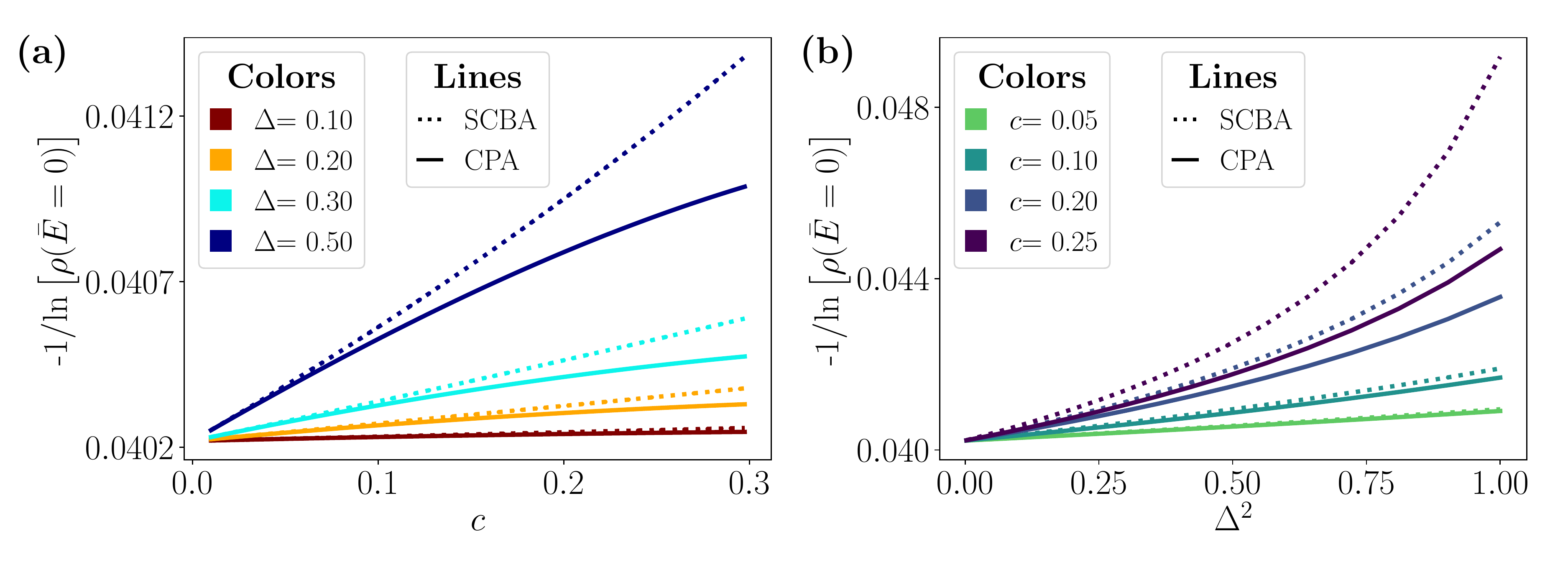}
    \caption{DOS for small disorder and low concentration of A impurities. The magnitude reported is the inverse of the logarithm of $\rho$ at zero energy as a function of $c$ and $\Delta^2$. We set $E_c=15$ in the numerical calculation. Energy and $\Delta$ are expressed in units of $\hbar v/a$.}
    \label{fig:CompRho}
\end{figure}

Figure~\ref{fig:CompRho} shows in more detail the range of equivalence of both approximations. For small $c$ and $\Delta$ ($c\lesssim 0.2$ and $\Delta \lesssim 0.5$ in the figure) the SCBA and CPA coincide whereas for higher values of disorder the overestimation of the SCBA becomes noticeable. Notice that, in the range of weak disorder, the analytic limit given by equation~\eqref{eq:small:rho_E0} is accurate and the DOS follows the exponential trend $-1/\ln \left[ \rho(\Et) \right] \sim c \Delta^2$. In a wider range of magnitude of disorder $\Delta$ and concentration $c$, the disagreement becomes apparent, leading to an excess of the DOS of the order of the value itself, as seen in figure~\ref{fig:EzeroPhases}. Moreover, the SCBA predicts a threshold for non-zero DOS smaller than the one predicted by the CPA, as shown in the figure~\ref{fig:EzeroPhases} (d), where the DOS is plotted as a function of $\Delta$. 

\begin{figure}[htb]
    \centering
    \includegraphics[width=\linewidth]{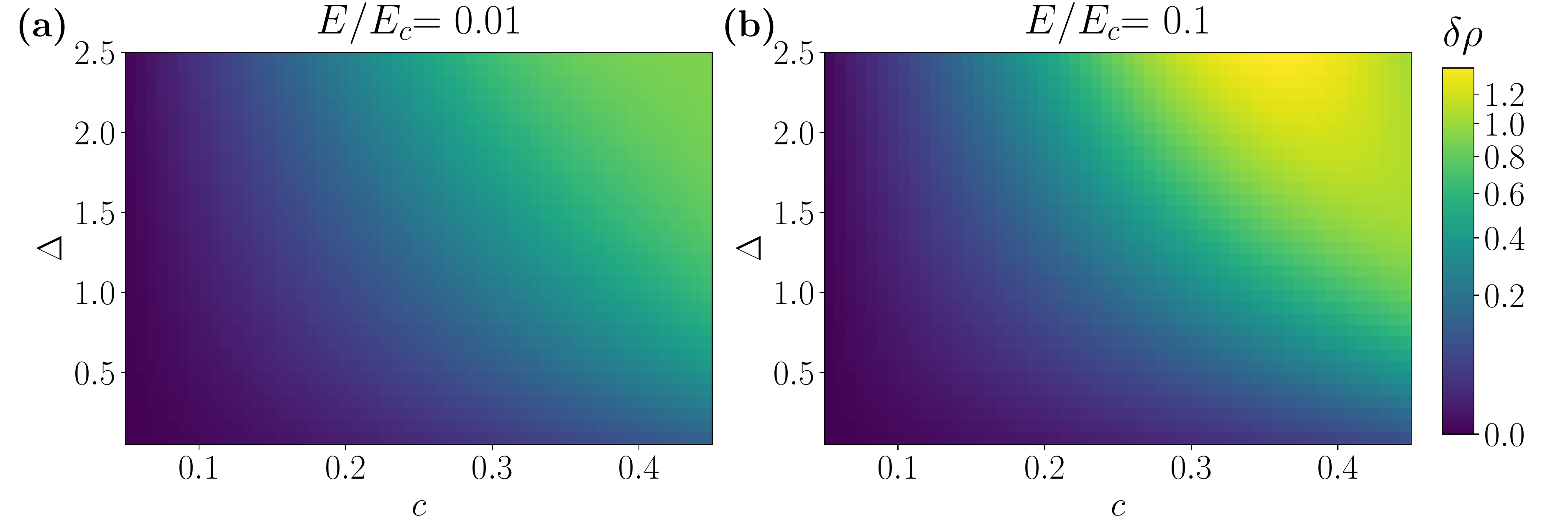}
    \caption{Relative error for the DOS at two different energies for $E_c=15$. Energy and $\Delta$ are expressed in units of $\hbar v/a$.}
    \label{fig:RelativeError}
\end{figure}

For non-zero energy, the tendency remains the same and the SCBA results in an overvaluation of the effect of the impurities. Due to the strictly non-zero DOS for $\abs{\Et}>0$, we can compute the relative error defined as
\begin{equation}
    \delta \rho (\Et) = 2\,\frac{\rho_\mathrm{SCBA} (\Et) - \rho_\mathrm{CPA} (\Et) }%
    {\rho_\mathrm{SCBA} (\Et) + \rho_\mathrm{CPA} (\Et) }~.
\end{equation}
Figure~\ref{fig:RelativeError} shows the relative error at a given energy as a function of $\Delta$ and $c$. We observe that the discordance grows with the energy, as  previously (see figure~\ref{fig:AnalyticsComp}).

We conclude by stressing the range of validity of the CPA. The CPA has been proven to reliably obtain the self-energy for a wide range of scenarios. It yields the correct result in the weak scattering limit (where it coincides with the SCBA), in the strong limit and in the dilute limit~\cite{Bruus04, Elliot74}. In fact, the only approximation assumed in the CPA condition is that if the averaged single-site $t-$matrix is zero [equation \eqref{eq:CPA:cond}], then the averaged $T-$matrix of the whole system is zero. This approximation is correct whenever the spatial correlation of disorder is negligible. The single-site CPA incorrectly treats multiple scattering terms associated with clusters of fixed number of neighbour sites~\cite{Economou06}. Diagrammatically, it corresponds to the fact that the self-energy in CPA does not include wigwam diagrams with crossing lines, whose contribution is negligible as long as the scattering length of the impurity potential is smaller than $a$~\cite{Bruus04}. Therefore, if the impurities are diluted and short-range order is absent, the results of the CPA are essentially exact.

Finally, let us stress the validity of the results obtained in this work for the understanding of other 2D Dirac materials. After a trivial rotation, the electron Hamiltonian~\eqref{eq:01a} is basically the same that of a low-energy electron in graphene. Hence, our results are of interest in the description of graphene impurities~\cite{Skrypnyk2018} specially in the non-magnetic impurities case. Starting from the seminal work by Noro \emph{et al.}~\cite{Ando1998}, graphene disordered sheets have been studied extensively within the SCBA approach~\cite{Hu2008, Noro2010}, showing a sizable effect of the disorder present in the samples. The numerical findings also show the behavior presented here for the DOS~\cite{Wu2008}. Moreover, proposals have been made in order to obtain the averaged DOS of those systems by measuring the quantum capacitance~\cite{Li2013}. 

\section{Conclusions} \label{sec:conclusions}

We have solved the effective medium approximation for a many-impurity scattering problem on a 2D surface of a topological insulator within the SCBA and CPA. Moreover, we have analysed in detail the differences, weaknesses and strengths of both methods. The simplicity of the SCBA allows us to extend the analytic calculations, bringing almost exact analytic results for small magnitude of  disorder and low concentration of impurities without the need for the numerical solution of the self-consistency conditions. On the other hand, the CPA enables us to exactly solve the problem for any number of single-impurity scattering events, yielding reliable results even in the non-pertubative limit. Moreover, as expected by the correspondence of SCBA and CPA for weak disorder, both approximations coincide in the range of dilute and weakly-interacting impurities. 

A reliable determination of the effective coupling constant, or equivalently, the self-energy, is of central importance since it allows us to calculate all physically meaningful quantities. Aiming to achieve this, it is crucial to use the appropriate method matching the regime of concentration and disorder strength properly. 
In conclusion, our finding thus not only calls for a revision of current theories based on the SCBA, but also provides a reliable implementation of the (more accurate) CPA for studying impurity scattering of 2D Dirac matter.

\appendix

\section{One-band approximation} \label{app:one-band}

Starting from equation~(\ref{eq:12}), the Green's function operators associated to $\widehat{H}_\mathrm{eff}$ and $\widehat{H}_0$ satisfy~\cite{Economou06}
\begin{equation}
  \widehat{G}_\mathrm{eff}=\widehat{G}_0+\widehat{G}_0\sum_{n}
  \mid\omega_n \,\rangle \lambda_\mathrm{CPA}(z) \langle\, \omega_n\mid\widehat{G}_\mathrm{eff}\ .
  \label{eq:01A}
\end{equation}
We now take into account the closure relation of the plane waves
\begin{equation}
  \sum_{\bm k} \mid {\bm k}\,\rangle \langle \,{\bm k}\mid=\mathbb{1}\ ,
  \label{eq:02A}
\end{equation}
$\mathbb{1}$ being the identity operator, and equation~(\ref{eq:05a}) to obtain
\begin{align}
  \langle\,{\bm k} \mid \widehat{G}_\mathrm{eff}\mid{\bm k}^{\prime}\rangle &= G_0({\bm k},z)
  \delta_{{\bm k},{\bm k}^{\prime}}+\frac{\lambda_\mathrm{CPA}(z)}{a^2}\,G_0({\bm k},z)\,\omega({\bm k})\nonumber \\
  &\times \sum_{\bm K} \omega^{*}({\bm k}+{\bm K})\langle\,{\bm k}+{\bm K}\mid \widehat{G}_\mathrm{eff}\mid{\bm k}^{\prime}\,\rangle\ .
  \label{eq:03A}
\end{align}
where the index ${\bm K}$ runs over the vectors of the reciprocal lattice of the impurity lattice. In the one-band approximation, the Fourier transform of the shape function is assumed to vanish outside the Brillouin zone~\cite{Sievert73,Glasser75}. In this way, we only retain the term ${\bm K}=0$ in the expansion~(\ref{eq:03A}). Therefore
\begin{align}
  \langle\,{\bm k} \mid \widehat{G}_\mathrm{eff}\mid{\bm k}^{\prime}\rangle &= 
  \left[
  1-\frac{\lambda_\mathrm{CPA}(z)}{a^2}\,G_0({\bm k},z)|\omega({\bm k})|^2
  \right]^{-1} G_0({\bm k},z) \delta_{{\bm k},{\bm k}^{\prime}} \ .
  \label{eq:04A}
\end{align}
The translational invariance of the effective medium ensures that the Green's function operator is diagonal in the basis of plane waves. The general relation between operators $(A-B)^{-1}=A^{-1}B(A-B)^{-1}$ allows us to rewrite~(\ref{eq:04A}) as
\begin{equation}
  \langle\,{\bm k} \mid \widehat{G}_\mathrm{eff}\mid{\bm k}^{\prime}\rangle=G_0\big[{\bm k},z-\Sigma_\mathrm{CPA}({\bm k},z)\big]
  \,\delta_{{\bm k},{\bm k}^{\prime}}\ ,
  \label{eq:05Aa}
\end{equation}
where $\Sigma_\mathrm{CPA}({\bm k},z)=\lambda_\mathrm{CPA}(z)|\omega({\bm k})|^2/a^2$.

Using the closure relation~(\ref{eq:02A}) we get
\begin{gather}
  \langle\, \omega_n \mid \widehat{G}_\mathrm{eff}(z)\mid\omega_n\,\rangle = \sum_{\bm k}\langle\,{\bm k} \mid
  \widehat{G}_\mathrm{eff}\mid{\bm k}\rangle\, |\langle\,{\bm k}\mid \omega_n\,\rangle|^2 =\frac{1}{S} \sum_{\bm k}\langle\,{\bm k} \mid \widehat{G}_\mathrm{eff}\mid{\bm k}\rangle\, |\omega({\bm k})|^2\ ,
  \label{eq:06A}
\end{gather}
where $S$ is the area of the system. After converting the sum over ${\bm k}$ into an integration we finally obtain~(\ref{eq:16b}).

\section{Calculation of the coupling constant in the SCBA} \label{app:SCBA:analytics}

As mentioned in the text, the SCBA self-consistent equation can be solved exactly at $\Et=0$ and approximately in the case of weak disorder. In the case of $\Et=0$, considering the symmetry properties of equation~\eqref{eq:selfSCBA:beta} in the main text, we find that the real part of the coupling constant must be zero. Therefore, replacing $\Lambda \to -i \Gamma$, we conclude that the self-consistent condition reduces to
\begin{equation}
    -\frac{1}{\beta} = \ln \left[ \frac{\Ga_\mathrm{SCBA}^2 (\Et = 0) }{E_c^2+\Ga_\mathrm{SCBA}^2 (\Et = 0) }\right]~,
\end{equation}
whose solution is given by equation~\eqref{eq:SCBA:G_E0}. 

In the weak disorder regime and for energies $\Et \ll E_c$, the coupling constant fulfils $\abs{\Lambda_\mathrm{SCBA}}(\Et) \ll E_c$. Therefore, we can expand the SCBA equation as
\begin{equation}
    \Lambda_\mathrm{SCBA}\simeq \beta (\Et-\Lambda_\mathrm{SCBA}) \ln\left[- \frac{(\Et- \Lambda_\mathrm{SCBA})^2}{E_c^2}\right]~,
\end{equation}
where $\Lambda_\mathrm{SCBA}=\Lambda_\mathrm{SCBA}(\Et)$. Considering solutions with $\IM(\Lambda_\mathrm{SCBA})<0$, we obtain equation~\eqref{Lambert}. This approximate solution resembles the exact case at $\Et=0$ for small $\beta$. In fact, at zero energy, we obtain equation~\eqref{expansion-Lambert}, which corresponds to the first term in the series expansion of equation \eqref{eq:SCBA:G_E0} for $\beta \ll 1$.

\section{Expression for the DOS }  \label{app:DOS}

The DOS per unit area is obtained from the Green's function using equation~\eqref{eq:rho:def}. In the case of the 2D effective Hamiltonian we are dealing with, this expression is written as 
\begin{equation} \label{eq:rho}
    \rho(\Et) = -\frac{1}{\pi}\IM \sum_{\tau=\pm1}
    \int\frac{\mathrm{d}^2{\bm k}}{4\pi^2}\,\frac{1}
    {\Et + i0^+-\tau \hbar v k-\Lambda(\Et)|{w(k)}|^2/a^2}~.
\end{equation} 
After some algebra and expressing the energy in units of $\hbar v_F/a$ and the coupling constant $\Lambda (\Et) $ as given by equation~\eqref{eq:Lambda}, when $\omega({\bm k})=a\theta(k_\mathrm{c}-k)$ we obtain the following expression. For the sake of simplify, hereafter we omit the dependence on $\Et$ in $\alpha \equiv \alpha(\Et) $ and $\Gamma \equiv \Gamma(\Et) $.
\begin{subequations}
\begin{align}
    \rho(\Et) & = \frac{1}{4\pi^2}\Bigg\{\Ga\ln\big[M(\Et)\big] 
    + 2(\Et-\alpha)\bigg[ \arctan\left(\frac{E_c+\alpha-\Et}{\Ga}\right) - \arctan \left(\frac{E_c-\alpha+\Et}{\Ga}\right) \nonumber \\ & +2\arctan\left(\frac{\Et-\alpha}{\Ga}\right)\Bigg] \Bigg\} +\frac{1}{2\pi}\Et\big[\theta(\Et-E_c) + \theta(\Et + E_c)-1\big]~,
    \label{eq:D2a}
\end{align}
where 
\begin{equation}
  M(\Et)  = \frac{[( E_c-\alpha+\Et)^2 +\Ga^2]\left[ (E_c+\alpha-\Et)^2+ \Ga^2\right]}{\left[\Ga^2+(\alpha-\Et)^2\right]^2}~.
\end{equation}
\end{subequations}

\ack

The authors thank A.\ D\'{\i}az-Fern\'{a}ndez 
for a critical reading of the manuscript. This work has been supported by Ministerio de Cien\-cia e In\-novaci\'{o}n (Grants MAT2016-75955 and PID2019-106820RB-C21).

\section*{References}

\bibliographystyle{iopart-num.bst}
\providecommand{\newblock}{}

\end{document}